\documentclass[prd,tightenlines,nofootinbib,showpacs,preprintnumbers,superscriptaddress, twocolumn]{revtex4-1}
\usepackage{amsfonts,amsmath,amssymb,amsthm,bbm,hyperref}
\usepackage{graphicx}
\usepackage{color}
\usepackage[T1]{fontenc}

\newcommand{\be}{\begin{equation}}
\newcommand{\ee}{\end{equation}}
\newcommand{\beq}{\begin{eqnarray}}
\newcommand{\eeq}{\end{eqnarray}}
\newcommand{\nn}{\nonumber}

\DeclareMathOperator\cotanh{cotanh}

\begin{document}

\title{Cosmological perturbations in the entangled inflationary universe}
\author{Salvador J. Robles-P\'{e}rez}
\affiliation{Estaci\'{o}n Ecol\'{o}gica de Biocosmolog\'{\i}a, Pedro de Alvarado 14, 06411 Medell\'{\i}n, Spain.}
\affiliation{Instituto de  F\'{\i}sica Fundamental, CSIC, Serrano 121, 28006 Madrid, Spain}
\date{\today}

\begin{abstract}
In this paper it is presented the model of a multiverse made up of universes which are created in entangled pairs that conserve the total momentum conjugated to the scale factor. For the background spacetime it is assumed a FRW metric with a scalar field with mass $m$ minimally coupled to gravity. For the fields that propagate in the entangled spacetimes it is considered the perturbations of the spacetime and the scalar field, whose quantum states become entangled too. They turn out to be in a quasi thermal state and the corresponding thermodynamical magnitudes are computed. Three observables are expected to be caused by the creation of the universes in entangled pairs: a modification of the Friedmann equation because the entanglement of the spacetimes, a modification of the effective value of the potential of the scalar field by the backreaction of the perturbation modes, and a modification of the spectrum of fluctuations because the thermal distribution induced by the entanglement of the partner universes. The later would be  a distinctive feature of the creation of universes in entangled pairs.
\end{abstract}

\maketitle



\section{Introduction}

One of the most beautiful features of quantum cosmology is the appearance of the classical spacetime and the quantum mechanics of  matter fields from the quantum state of the universe. From the point of view of quantum cosmology these are emergent features of the semiclassical regime of the wave function of the universe \cite{Hartle1990}. This is not very surprising because the wave function of the universe is obtained by quantizing the Einstein-Hilbert action and the action of the matter fields so, in a top-down approach one must recover in the appropriate limit the (semi-)classical behaviour of the spacetime and the matter fields.

The wave function of the spacetime and the matter fields, all together\footnote{In the context of a single universe $\Psi$ is called the \emph{wave function of the universe} \cite{Hartle1983}. However, that name can be misleading in the context of the multiverse so we here retain the name wave function of the spacetime and matter fields for the wave function that describes the quantum state of the whole spacetime manifold and the matter fields that  propagate therein.}, $\Psi$, can be obtained by solving the Hamiltonian constraint 
\be\label{HC01}
\hat{H} \Psi = 0 ,
\ee
where $\hat{H}$ is the operator form of the Hamiltonian associated to the total action. The Hamiltonian constraint (\ref{HC01}) turns out to be a very complicated equation. However, for most of the evolution of the universe this is described by a homogeneous and isotropic background with small energy fields propagating therein. In that case, the Hamiltonian constraint (\ref{HC01}) can be re-written as
\be\label{HC02}
 (\hat{H}_{bg} + \hat{H}_m) \Psi = 0 ,
\ee
where $H_{bg}$ is the Hamiltonian of the background spacetime and $H_m$ contains the matter degrees of freedom. The wave function, $\Psi = \Psi(q_{bg}, q_{m})$, where $q_{bg}$ are the degrees of freedom of the background and $q_m$ are the matter degrees of freedom, can then be expressed in the semiclassical regime as a linear combination of WKB solutions, i.e. \cite{Hartle1990}
\be\label{SCWF01}
\Psi(q_{bg}, q_{m}) = \sum C(q_{bg}) e^{\pm \frac{i}{\hbar} S_0(q_{bg})} \chi(q_{bg}, q_{m}) ,
\ee
where $C(q_{bg})$ is a slow-varying function of the background variables, $S_0(q_{bg})$ is the action of the background spacetime, and $\chi(q_{bg}, q_{m})$ is the wave function of the inhomogeneous degrees of freedom that propagate in the homogeneous and isotropic background. Inserting the semiclassical wave function (\ref{SCWF01}) into the Hamiltonian constraint (\ref{HC02}) and solving it order by order of $\hbar$ in $H_{bg}$, it is obtained: at zero order, the classical equations of the background spacetime, which provide us with the time variable for the fields that propagate therein and, at first order in $\hbar$, it is recovered the Schr\"{o}dinger equation of the matter fields with the time variable of the background spacetime provided by the zero order equations. Therefore, all the physics we know can in principle be derived from a semiclassical state like (\ref{SCWF01}).

Each single addend in (\ref{SCWF01}) can quantum mechanically represent the state of a spacetime background with matter fields propagating therein. The semiclassical state (\ref{SCWF01}) should be seen then as the quantum state of a many universe system. However, once the decoherence process has taken placed between the branches in (\ref{SCWF01}) \cite{Halliwell1989, Kiefer1992}, the customary approach consists of considering one of these branches as the representative of our universe and disregard the rest of them as being physically redundant or just account them for statistical measures.  One can still consider other possibilities. For instance, one can look for non-local interactions or quantum correlations between the  branches in (\ref{SCWF01}) ultimately rooted in a common origin or derived from residual interacting terms of the underlying theory, whether this is one of the string theories or the quantum theory of gravity. In any of these cases, one cannot disregard the rest of branches of the general state (\ref{SCWF01}).

In particular, in this paper we are going to study the effects that the creation of universes in entangled pairs may have in the properties of each single universe of the entangled pair. Let us notice that the semiclassical state (\ref{SCWF01}) can be rearranged as
\be\label{SCWF02}
\Psi =  \sum C e^{ \frac{i}{\hbar} S_0} \chi + C^* e^{- \frac{i}{\hbar} S_0} \chi^*  .
\ee
Each term in the sum (\ref{SCWF02}) can be seen as the wave function of a pair of entangled branches or universes. We shall shown that each branch of the entangled pair has an opposite momentum conjugated to the background variables and thus, the creation of universes in entangled pairs conserve the total momentum in a parallel way as the creation of particles in entangled pairs conserve the total momentum in a quantum field theory \cite{RP2017c}. The creation of universes in entangled pairs would have important consequences in the properties of the matter fields that propagate in their spacetimes because the quantum state of the fields of the two universes become entangled too with a rate of entanglement that depends on the entanglement properties of the parent universes. Then, the effects of the creation of universes in entangled pairs are encoded in the properties of the matter fields that propagate in each single universe, and thus, they might be observed in the properties of our universe \cite{RP2017c}.

In this paper we provide the detailed model of a multiverse made up of  universes which are created in entangled pairs. It comes to complement the model presented in Ref. \cite{RP2017c}, where it is considered a conformally coupled massless scalar field propagating in a homogeneous and isotropic background spacetime. It provides analytical solutions for the quantum state of the universes and a clear picture of the entanglement processes that may occur in the multiverse. We here consider a minimally coupled scalar field with mass $m$ that can mimic more accurately the early stage of our universe. The paper is outline as follows. In Sect. II it is obtained the dynamics of the background spacetime and the Schr\"{o}dinger equation of the matter fields from the semiclassical state of the spacetime and the matter fields. In Sect. III it is shown that the most natural way in which the universes can be created is in entangled pairs that conserve the total momentum. In Sect. IV it is imposed the boundary condition that the perturbation modes are in the composite vacuum state of the invariant representation that represents a stable vacuum state along the entire evolution of the field. However, in terms of the instantaneously diagonal representation, the invariant vacuum state is full of particle-antiparticle pairs, being the former created in the observer's universe and the antiparticle in the partner one. In Sect. V it is computed the quantum state of the particles in each single universe of the entangled pair and the  the thermodynamical magnitudes of the field. In Sect. VI we present three observables that are expected to be caused by the creation of universes in entangled pairs. Finally, in Sect. VII we shall summarize and draw some conclusions.

\section{Spacetime background and the quantum mechanics of  matter fields}

For the background of the model of our universe let us consider a homogeneous and isotropic spacetime with FRW metric 
\be\label{G01}
ds^2 = - dt^2 + a^2(t) d\Omega^2_3  ,
\ee
where $d\Omega_3^2$ is the line element on the unit three sphere, and a homogeneous and isotropic scalar field, $\phi(t)$, minimally coupled to gravity, with mass $m$ given  by
\be
m = V''(\phi) ,
\ee
where $V(\phi)$ is the potential of the scalar field. We shall leave unfixed the functional form of the potential in order to potentially consider different cases including convex ($V'' > 0$) as well as concave ($V'' < 0$) potentials.

For the fields that propagate in the homogeneous and isotropic background we shall consider the small perturbations of the spacetime and the scalar field, i.e. the gravitons and the field particles. These are going to be described by two fields \cite{Halliwell1985, Kiefer1987}
\beq\label{hdecomp}
h_{ij}(t,\textbf{x}) - a^2 \Omega_{ij} &=& a^2  \sum_\textbf{n} 2 d_\textbf{n}(t) G^\textbf{n}_{ij}(\textbf{x}) + \ldots  , \\ \label{fidecomp}
\phi(t,\textbf{x}) - \frac{1}{\sqrt{2\pi}} \phi(t) &=& \sum_\textbf{n} f_\textbf{n}(t) Q^\textbf{n}(\textbf{x}) ,
\eeq
where $Q^\textbf{n}(\textbf{x})$ are the scalar harmonics on the three-sphere and $G^\textbf{n}_{ij}(\textbf{x})$ the transverse traceless tensor harmonics \cite{Halliwell1985}, with, $\textbf{n}\equiv(n,l,m)$. We shall only focus on the scalar modes of the perturbed field, $f_\textbf{n}$, and the tensor modes of the perturbed spacetime, $d_\textbf{n}$, as representative examples of the matter particles and the gravitons, respectively.

The background degrees of freedom are therefore the scale factor, $a$, and the homogeneous and isotropic part of the scalar field, $\phi$, and the \emph{matter} degrees of freedom are the perturbation modes $f_\textbf{n}$ and $d_\textbf{n}$, denoted generically by $x_\textbf{n}$. The semiclassical wave function of the  spacetime and the matter fields, $\Psi(a,\phi;x_\textbf{n})$, is then given by a composite state of the wave function that represents the quantum state of the homogenous and isotropic background, $\Psi_0(a,\phi)$, and a wave function that contains the  degrees of freedom of the perturbation,
\be\label{WF01}
\Psi(a,\phi;x_\textbf{n}) = \Psi_0(a,\phi) \chi(a,\phi; x_\textbf{n}) .
\ee 
The wave function $\Psi_0$ is the solution of the Wheeler-DeWitt equation of the homogeneous and isotropic background, 
\be\label{WDW01}
\hat{H}_0 \Psi_0 = 0 ,
\ee
where
\be\label{H0}
\hat{H}_0 =  \frac{1}{2 a} \left( \frac{\partial^2}{\partial a^2} + \frac{1}{a} \frac{\partial}{\partial a} - \frac{1}{a^2} \frac{\partial^2}{\partial \phi^2} + a^4 H^2(\phi) - a^2  \right)  ,
\ee
with, $H^2(\phi) \equiv 2 V(\phi)$. In the semiclassical regime we can consider the WKB solutions of the Wheeler-DeWitt equation (\ref{WDW01}), 
\be\label{SCWF03}
\Psi_0^\pm(a,\phi) = C(a,\phi) e^{\pm\frac{i}{\hbar} S(a,\phi)} .
\ee
Inserting the wave function (\ref{SCWF03}) into the Hamiltonian constraint (\ref{H0}) it is satisfied, at zero order in $\hbar$, the classical Hamilton-Jacobi equation \cite{Kiefer1987}
\be\label{HJ01}
-\left( \frac{\partial S}{\partial a} \right)^2 +\frac{1}{a^2} \left( \frac{\partial S}{\partial \phi} \right)^2 + a^4 H^2(\phi) - a^2 = 0 .
\ee
Then, a WKB-time parameter $t$ can be defined \cite{Kiefer1987} 
\be\label{WKBt01}
\frac{\partial}{\partial t} = \pm\nabla S \cdot \nabla \equiv \pm\left( -\frac{1}{a} \frac{\partial S}{\partial a}\frac{\partial }{\partial a} +\frac{1}{a^3} \frac{\partial S}{\partial \phi}\frac{\partial }{\partial \phi} \right) ,
\ee
where $\nabla$ is the gradient of the minisuperspace \cite{Halliwell1990}. In terms of the WKB time (\ref{WKBt01}),
\be\label{MOM01}
\dot{a}^2 = \frac{1}{a^2} \left( \frac{\partial S}{\partial a}\right)^2 \ , \ \dot{\phi}^2 = \frac{1}{a^6} \left( \frac{\partial S}{\partial \phi} \right)^2 ,
\ee
and the Hamilton-Jacobi equation (\ref{HJ01}) turns out to be the Friedmann equation of the background spacetime
\be\label{FE01}
\dot{a}^2 + 1 - a^2 \left( \dot{\phi}^2 + 2 V(\phi) \right) = 0 .
\ee
On the other hand, inserting the wave function (\ref{WF01}) into the total Hamiltonian, $H = H_0 + H_m$, where $H_m$ is the Hamiltonian of the perturbation modes, it is obtained at first order in $\hbar$ of $H_{0}$, 
\be\label{SCH00}
\mp i \hbar \left( -\frac{1}{a} \frac{\partial S}{\partial a}\frac{\partial }{\partial a} +\frac{1}{a^3} \frac{\partial S}{\partial \phi}\frac{\partial }{\partial \phi} \right) \chi = H_m \chi ,
\ee
which is the Schr\"{o}dinger equation of the matter fields that propagate in the background spacetime (\ref{G01}) provided that the time variable of the Schr\"{o}dinger equation is the WKB time defined in (\ref{WKBt01}) with the positive sign for the semiclassical wave function $\Psi^-$ in (\ref{SCWF03}) and the negative sign for $\Psi^+$. The wave function (\ref{WF01}) can then be written as
\be\label{SCWF04}
\Psi = C e^{+\frac{i}{\hbar} S} \chi_+ + C^* e^{-\frac{i}{\hbar} S} \chi_- ,
\ee
with, $\chi_- = \chi_+^*$, satisfying the Schr\"{o}dinger equation (\ref{SCH00}), i.e.
\be\label{SCH01}
i \hbar \frac{\partial }{\partial t_{\pm}} \chi_{\pm} = H_m \chi_{\pm} ,
\ee
where, $\chi_\pm(t,x_\textbf{n}) \equiv \chi_\pm(a, \phi; x_\textbf{n})$, evaluated at the  background solutions, $a(t)$ and $\phi(t)$ of the Friedmann equation (\ref{FE01}). Let us notice that the Friedmann equation is invariant under the time reversal symmetry, $t \leftrightarrow - t$. 

If we restrict to small linear perturbations the different modes do not interact \cite{Halliwell1987, Kiefer1992} and $H_m$ turns out to be
\be
H_m =  \sum_\textbf{n} H_\textbf{n} ,
\ee 
with
\be\label{Hn}
H_\textbf{n} = \frac{1}{2 M} p_{x_\textbf{n}}^2 + \frac{M \omega_n^2}{2} x_\textbf{n}^2 ,
\ee
where, $M(t) = a^3(t)$, and \cite{Kiefer1987}
\be\label{OMEGAM01}
\omega_n^2 = \frac{n^2-1}{a^2} ,
\ee
for the tensorial modes of the spacetime ($x_\textbf{n} \equiv d_\textbf{n}$), and 
\be\label{OMEGAM02}
\omega_n^2 = \frac{n^2-1}{a^2} \pm m^2    ,
\ee
for the perturbation modes of the scalar field  ($x_\textbf{n} \equiv f_\textbf{n}$). In (\ref{OMEGAM02}) , the $+$ sign corresponds to the oscillatory phase and the $-$ sign to the inflationary stage of the scalar field $\phi$ \cite{Kiefer1987}. The perturbation modes satisfy then the wave equation of the harmonic oscillator
\be\label{HOEQ01}
\ddot{x}_\textbf{n} + \frac{\dot{M}}{M} \dot{x}_\textbf{n} + \omega_n^2 x_\textbf{n} = 0 .
\ee

The wave function of the perturbation modes can then be written as \cite{Kiefer1987, Grishchuk1990, Kiefer1992}
\be\label{CHI01}
\chi = \prod_\textbf{n} \chi_\textbf{n}(t, x_\textbf{n}) ,
\ee
where the function $\chi_\textbf{n}(t,x_\textbf{n})$ is the wave function of a harmonic oscillator with time dependent mass and frequency, whose general solution can be expanded in the basis of number eigenstates of the invariant representation, $\psi_{N,\textbf{n}}$, as
\be
\chi_\textbf{n} = \sum_N C_N \psi_{N,\textbf{n}} ,
\ee
where $C_N$ are constants coefficients and the wave function of the invariant number state, $\psi_{N,\textbf{n}}$, is given by \cite{Leach1983, RP2017f}
\beq\nn
\psi_{N,n}(a,\phi; x_\textbf{n}) &\equiv& \langle a, \phi; x_\textbf{n} | N_\textbf{n} \rangle = \\ \label{NS01}
&=& \frac{1}{\sigma^\frac{1}{2}} \exp\left\{ \frac{i M}{2}\frac{\dot{\sigma}}{\sigma} x_\textbf{n}^2 \right\} \bar{\psi}_N\left(\frac{x_\textbf{n}}{\sigma}, \tau \right) 
\eeq
where $\bar{\psi}_N(q,\tau)$, with $q\equiv \frac{x_\textbf{n}}{\sigma}$, is the customary wave function of the harmonic oscillator, i.e.
\be
\bar{\psi}_N(q,\tau) = \left( \frac{1}{2^N N! \pi^\frac{1}{2} } \right)^\frac{1}{2} e^{-i (N+\frac{1}{2}) \tau} e^{-\frac{q^2}{2}}  {\rm H}_N(q) ,
\ee
with ${\rm H}_N$ being the Hermite polynomial of degree $N$, and $\tau = \tau(t)$ is given by
\be\label{TAU01}
\tau(t) = \int^t \frac{1}{M(t') \sigma^2(t')} dt' ,
\ee
and $\sigma(t)$ is an auxiliary function that satisfies the non-linear equation \cite{Lewis1969, Leach1983}
\be\label{SIGM01}
\ddot{\sigma} + \frac{\dot{M}}{M} \dot{\sigma} + \omega_n^2 \sigma = \frac{1}{M^2 \sigma^3} .
\ee
It is worth noticing that a solution of (\ref{SIGM01}) can generally be written as
\be\label{SIGM02}
\sigma = \sqrt{\sigma_1^2 + \sigma_2^2 } ,
\ee
where $\sigma_1$ and $\sigma_2$ are two independent solutions of (\ref{HOEQ01}) satisfying some specific boundary condition. For the boundary condition one has to realized that in terms of the variable, $z_\textbf{n} \equiv a x_\textbf{n}$, in conformal time $\eta = \int \frac{dt}{a}$, the equation of the harmonic oscillator (\ref{HOEQ01}) turns out to be
\be
z''_\textbf{n} + n^2 z_\textbf{n} = 0 ,
\ee 
in the limit of large modes. In that limit, and in terms of the variable $z_\textbf{n}$, the wave function of the modes should be then the customary wave function of the harmonic oscillator with unit mass and constant frequency $n$. This is accomplished if we impose that 
\be\label{SIGMBC01}
\sigma \rightarrow \frac{1}{\sqrt{M \omega_n }} \approx \frac{1}{a\sqrt{n}} \ , \ \dot{\sigma} \rightarrow 0 ,
\ee
 in the limit of large modes for all time. Thus, the computation of the wave function of the perturbation modes essentially reduces to the computation of the solutions of (\ref{SIGM01}) that satisfy the boundary condition (\ref{SIGMBC01}). Their quantum state however, will depend on the boundary condition that we impose on the state of the field and this, in turn, will depend on the boundary condition imposed on the state of the universes. In particular, it will be shown that in the context of the creation of universes in entangled pairs it depends on the rate of entanglement between the universes.

\section{Creation of universes in entangled pairs}

During the early stage of the universe the potential of the field can be considered approximately constant for the tiny amount of time for which the universe is exponentially expanding. In that case, the wave function of the universe, $\Psi$, can be expanded in partial waves
\be\label{WF02}
\Psi(a,\phi) = \int \frac{d K}{\sqrt{2 \pi}} e^{ i K \phi} \Psi_K(a) \chi_K(a;x_\textbf{n}) ,
\ee
where the amplitude, $\Psi_K(a) \equiv \Psi_K(a,\phi_0)$, satisfies the Wheeler-DeWitt like equation
\be\label{WDW02}
 \frac{\partial^2{\Psi}_K}{\partial a^2} + \frac{1}{a} \frac{\partial{\Psi}_K}{\partial a} + \Omega_K^2(a) \Psi_K(a) = 0,
\ee
with, 
\be\label{OMEGAK}
\Omega_K = \sqrt{H^2 a^4 - a^2 + \frac{K^2}{a^2}} ,
\ee
where, $H^2 \equiv V(\phi_0)$, evaluated at some initial value $\phi_0$. The wave function of the modes, $\chi_K(a;x_\textbf{n}) \equiv \chi_K(a,\phi_0;x_\textbf{n})$, satisfies then the Schr\"{o}dinger equation (\ref{SCH01}) with the time variable $t$ of the corresponding  background spacetime.

However, the structure of the wave function $\Psi$ in (\ref{SCWF02}) and (\ref{SCWF04}) suggests that the universes might be created in entangled pairs (see also Refs. \cite{RP2011b, Garay2014, RP2014, RP2017c}). In that case, the perturbation modes of the scalar field and the gravitons of the spacetime  of each universe propagate in their corresponding background spacetimes, separately but in a quantum state that is correlated with the quantum state of the perturbations of the partner universe. Their quantum states become entangled too.

The creation of universes in entangled pairs can be easily visualized as the Lorentzian continuation of a double Euclidean instanton \cite{RP2014, Garay2014} (see, also, Ref. \cite{Bouhmadi2017}). However, it can be considered a more general feature in quantum cosmology. Let us notice that the mode decomposition (\ref{WF02}) resembles the mode decomposition customary used in a  quantum field theory. One can formally proceed as it is usually done in a quantum field theory and express the wave function (\ref{WF02}) in terms of two linearly independent solutions of the Wheeler-DeWitt equation (\ref{WDW02}). For these let us take the WKB solutions, given by
\be\label{WKB01}
\Psi_K^\pm \approx \frac{1}{\sqrt{a \, \Omega_K(a)}} e^{\pm i \int \Omega_K(a) d a } ,
\ee  
which are normalized according to, $\Psi_K^* \partial_a{\Psi}_K  - \Psi_K \partial_a{\Psi}^*_K = \pm \frac{2 i }{a}$. Then, (\ref{WF02}) can be written
\be\label{WF03}
\Psi(a,\phi) = \int \frac{dK}{\sqrt{2 \pi}} \left( e^{i K \phi} \Psi_K^+ \chi_K b_K + e^{-i K \phi} \Psi_K^- \chi^*_K c_K^* \right) , 
\ee
where $b_K$ and $c_K^*$ are two constants that are promoted to quantum operators, $b_K\rightarrow \hat{b}_K$ and $c_K^*\rightarrow\hat{c}^\dag_K$, in a quantum field theory of the wave function of the universe, the so-called third quantization formalism \cite{Strominger1990, RP2010}. The wave function (\ref{WF03}) describes then the quantum state of pairs of entangled universes that are created with opposite momenta, given by \cite{RP2017c, Garay2014}
\be\label{QMOM01}
\langle \hat{p}_a \rangle  \approx \pm \Omega_K ,
\ee
at leading order. The pair of newborn universes conserve thus the total momentum conjugated to the scale factor. The process parallels the creation of particles in entangled pairs with opposite momenta that conserve the total momentum in a quantum field theory described in an isotropic background spacetime. Therefore, in the context of the multiverse the creation of universes in entangled pairs seems to be the most natural way in which the universes can be created \cite{RP2017c}.

The momentum conjugated to the scale factor, however, depends on the expansion rate of the universe. The two WKB branches in (\ref{WKB01}) correspond then to an expanding and a contracting branch of the spacetime in terms of the time variable $t$. This can be seen by noticing that in the semiclassical regime the expected value of the momentum conjugated to the scale factor (\ref{QMOM01}) is highly peaked around the classical value, $p_a^c \equiv - a \dot{a}$ (see also (\ref{MOM01})). Thus, 
\be\label{FE03}
- a \dot{a} = \mp \Omega_K ,
\ee
which is nothing more than the Friedmann equation associated to the Wheeler-DeWitt equation (\ref{WDW02}), 
\be\label{FE04}
\frac{\dot{a}^2}{a^2} = \frac{\Omega_K^2}{a^4}  \Rightarrow \dot{a} = \pm \frac{\Omega_K}{a} ,
\ee
for the two branches $\Psi^\pm$. Nevertheless, the WKB time variable $t$ is not the time experienced by the internal observers in their particle physics experiments. Let us notice that the value of the momentum conjugated to the scale factor determines the value of the time variable in each single universe. Thus, according to (\ref{FE03}) the time variable in one of the universes of the entangled pair is defined by
\be\label{TI}
\frac{\partial }{\partial t_I} = \frac{\Omega_K}{a} \frac{\partial}{\partial a} ,
\ee 
and the time variable in the partner universe by
\be\label{TII}
\frac{\partial }{\partial t_{II}} = - \frac{\Omega_K}{a} \frac{\partial}{\partial a}  .
\ee
The time variables of the entangled universes are related by an antipodal like symmetry \cite{Linde1991}, $t_I = - t_{II}$, and the two branches turn out to be expanding branches in terms of the time variables $t_I$ and $t_{II}$. These are the time variables experienced by the internal observers in their particle physics experiments \cite{RP2017c, RP2017e}, which are governed by the Schr\"{o}dinger like equation
\be
\mp i \frac{\Omega_K}{a} \frac{\partial }{\partial a} \chi = H_m \chi .
\ee
This is the usual Schr\"{o}dinger equation with the time variable of each single universe of the entangled pair provided that the time variables in the two universes are taken to be $t_I$ and $t_{II}$ defined by (\ref{TI}) and (\ref{TII}), respectively. Therefore, $t_I$ and $t_{II}$ are the time variables measured by their actual clocks, and the two entangled universes become expanding universes from the point of view of the internal observers.

The wave function (\ref{WF03}) can now be written as 
\be\label{WF04}
\Psi = \int \frac{d K}{\sqrt{2\pi}}  \left( e^{i K \phi} \Psi^+_K \chi_K^I \hat{b}_K 
+  e^{- i K \phi} \Psi^-_K \chi_K^{II} \hat{c}_K^\dag \right),
\ee
where $\chi_K^I$ and $\chi_K^{II}$, with $\chi_K^{II} = (\chi_K^I)^*$, are the wave functions of the perturbation modes in each single universe of the entangled pair. They satisfy  the Schr\"{o}dinger equation (\ref{SCH01}) with $t_\pm$ being replaced by $t_I$ or by $t_{II}$ and $\chi_\pm$ by $\chi_K^I$ and $\chi_K^{II}$, respectively. In the case that the matter content of the universe is represented by a complex scalar field, then, matter would be always created in the observer's universe and antimatter in the partner universe, restoring thus the matter-antimatter symmetry in the pair of entangled universes \cite{RP2017e}.

\section{Quantum state of the perturbation modes}

In the context of the creation of universes in entangled pairs the matter particles and the gravitons of each universe propagate in the background spacetime of their corresponding universe and follow the Schr\"{o}dinger equation (\ref{SCH01}) with the time variable of the corresponding background spacetime. The wave function $\chi_K^{I,II}$ defines the quantization of the perturbation modes in the Schr\"{o}dinger picture, where $\hat{x}_\textbf{n}$ and $\hat{p}_{x_\textbf{n}}$ are time-independent operators that act on the time-dependent wave function $|\chi(t)\rangle$. In the configuration space of the amplitude of the perturbation modes the Hilbert space is spanned by the basis, $\{\prod_\textbf{n} |x_\textbf{n}\rangle\}$, where the vectors $|x_\textbf{n}\rangle$ are the eigenvectors of the amplitude operators $\hat{x}_\textbf{n}$. The general quantum state of the perturbations is then given by
\be
| \chi_K(t) \rangle = \int \prod_\textbf{n} dx_\textbf{n} \chi(x_\textbf{n}, t ) \prod_\textbf{n}  | x_\textbf{n} \rangle .
\ee
In the case of small perturbations, for which the different modes do not interact among them and $\chi(x_\textbf{n}, t )$ can be written as in (\ref{CHI01}), it can be written 
\be
| \chi_K(t) \rangle = \prod_\textbf{n} \int dx_\textbf{n}  \chi_\textbf{n}(x_\textbf{n} , t)  | x_\textbf{n}  \rangle ,
\ee
where, $\chi_\textbf{n}(x_\textbf{n}, t) = \langle x_\textbf{n} | \chi(t)\rangle$.

However, it seems to be more useful the development of the corresponding  quantum field theory. This can be done as usual, by considering in the wave equation of the the fields that represent the inhomogeneous degrees of freedom, given by (\ref{hdecomp}-\ref{fidecomp}), the general solution of the harmonic oscillator (\ref{HOEQ01}) with $\omega_n$ being given by (\ref{OMEGAM01}) in the case of the perturbation modes of the scalar field, and by (\ref{OMEGAM02}) in the case of the gravitons of the gravitational field. It can then be written
\be
x_\textbf{n}(t) = v_n^*(t) a_\textbf{n}^- + v_n(t) b_{-\textbf{n}}^+ ,
\ee
where $v_n(t)$ and $v_n^*(t)$ are two linearly independent solutions of the harmonic oscillator (\ref{HOEQ01}), and $a_\textbf{n}^-$ and $b_\textbf{n}^+$ are two constants satisfying, $b_\textbf{n}^+ = (a_\textbf{n}^-)^*$. Then, the development of the quantum field theory of the perturbation modes in the two entangled universes follows by promoting the constants $a_\textbf{n}^-$ and $b_\textbf{n}^+$ to quantum operators, $a_\textbf{n}^- \rightarrow\hat{a}_\textbf{n}$ and $b_\textbf{n}^+ \rightarrow \hat{b}_\textbf{n}^\dag$, satisfying the customary commutation relations. In the picture of a pair of universes created in an entangled state, the symmetry of the composite state (\ref{WF04}) suggests that matter is created in the observer's universe and  antimatter in the partner universe \cite{RP2017e}. Let us notice that for the observer of the partner universe it is the other way around. Therefore, $\hat{a}_\textbf{n}^\dag$ and $\hat{a}_\textbf{n}$ are the creation and annihilation operators of matter in one universe and, $\hat{b}_\textbf{n}^\dag$ and $\hat{b}_\textbf{n}$, are the creation and annihilation of matter in the other universe, satisfying both the corresponding commutation relations. In the case of a real field the particles are their own antiparticles. However, the creation and annihilation operators of modes of the two universes commute among them because to the Euclidean gap between the universes \cite{RP2014, RP2017e}, so we retain the different names $\hat{a}_\textbf{n}$ and $\hat{b}_\textbf{n}$. Then, the inhomogeneous part of the mode decomposition given in (\ref{hdecomp}-\ref{fidecomp}) can be written quantum mechanically as
\be
x(\textbf{x}, t) = \sum_\textbf{n} v_n^*(t) F_\textbf{n}(\textbf{x}) \hat{a}_\textbf{n} + v_n(t) F_{\textbf{n}}^* \hat{b}_\textbf{n}^\dag ,
\ee
where $F_n(\textbf{x})$ are the scalar harmonics on the three-sphere, $Q^\textbf{n}(\textbf{x})$, in the case of the perturbation modes of the scalar field, and $F_\textbf{n}$ are the transverse traceless tensor harmonics,  $G^\textbf{n}_{ij}(\textbf{x})$, in the case of the gravitons.

We can define the vacuum state of the particles in the two universes as , $|0\rangle_I$ and $|0\rangle_{II}$, as
\be
|0\rangle_{I,II} = \prod_\textbf{n} |0_\textbf{n}\rangle_{I,II} ,
\ee
where the states $|0_\textbf{n}\rangle$ are the states annihilated by $\hat{a}_\textbf{n}$ in the universe $I$, and by $\hat{b}_\textbf{n}$ in the universe $II$. On the other hand, with the operators $\hat{a}_\textbf{n}^\dag$ and $\hat{b}_\textbf{n}^\dag$ we can build the customary orthonormal bases for the corresponding Hilbert spaces,
\be
\prod_\textbf{n} | N_\textbf{n} \rangle_I = \prod_\textbf{n} \frac{1}{\sqrt{N_\textbf{n}!}} \left( \hat{a}_{n}^\dag \right)^{N_\textbf{n}} |0_\textbf{n}\rangle_I ,
\ee
and a similar one with $\hat{b}_{n}^\dag$ instead of $\hat{a}_{n}^\dag$ for the universe $II$. An arbitrary quantum state of the perturbations can then be written as a linear combination of the excited states
\be
|\chi \rangle_{I,II} = \sum_\textbf{n} \sum_{N_\textbf{n}} C_{N_\textbf{n}, N_{\textbf{n}'} \ldots} \prod_\textbf{n}  | N_\textbf{n} \rangle_{I,II}  ,
\ee
with constants coefficients, $C_{N_\textbf{n}, N_{\textbf{n}'} \ldots}$.

The quantum state of the perturbation modes are now determined by the boundary condition imposed on the states of the perturbation modes. This, in turn, depends on the boundary condition imposed on the state of the hosting universes. In the case of a pair of entangled universe, we impose that the perturbation modes are in the composite vacuum state of the invariant representation \cite{RP2017d}. The invariant representation has the great advantage that once the field is in a number state of the invariant representation it remains in the same state along the entire evolution of the field. In particular, once the field is in the vacuum sate of the invariant representation it remains in the same vacuum state along the entire evolution of the field. It seems to be then an appropriate boundary condition for the perturbation modes of the pair of entangled universes. The invariant representation of the harmonic oscillator (\ref{HOEQ01}) can be written as \cite{Lewis1969, RP2010}
\beq\label{IR01a}
\hat{a}_\textbf{n} &=& \sqrt{\frac{1}{2}} \left(  \frac{1}{\sigma} x_\textbf{n} + i ( \sigma p_{x_\textbf{n}}  - M \dot{\sigma} x_\textbf{n}  )  \right) , \\ \label{IR01b}
\hat{b}_{-\textbf{n}}^\dag &=& \sqrt{\frac{1}{2}} \left(  \frac{1}{\sigma} x_\textbf{n} - i ( \sigma p_{x_\textbf{n}}  - M \dot{\sigma} x_\textbf{n} )  \right) .
\eeq
The perturbation modes are then in the vacuum state of the invariant representation given by
\be\label{VS01}
|0 \rangle = |0_a 0_b \rangle = |0_a\rangle_I  |0_b\rangle_{II} .
\ee
However, the particles of the scalar field and the gravitons measured by the internal observers of the universe are not described by the number states of the invariant representations (\ref{IR01a}-\ref{IR01b}). They are instead described by the number states of the instantaneous diagonal representation of the Hamiltonian of the harmonic oscillator (\ref{Hn}), which defines the instantaneous vacuum state at each moment of time. In terms of the diagonal representation the amplitude of the perturbations and their conjugate momenta can be written as
\beq\label{DR01a}
x_\textbf{n} &=& \frac{1}{\sqrt{2 M \omega_n}} \left( {c}_\textbf{n} + {d}_{-\textbf{n}}^\dag \right) , \\ \label{DR01b}
p_{x_\textbf{n}} &=& - i \sqrt{\frac{M \omega_n}{2}} \left( {c}_\textbf{n} - {d}_{-\textbf{n}}^\dag \right) .
\eeq 
The invariant representation (\ref{IR01a}-\ref{IR01b}) can be related to the diagonal representation (\ref{DR01a}-\ref{DR01b}) through the Bogolyubov transformation
\beq\label{BT01a}
{a}_\textbf{n} &=& \mu(t) \, {c}_{\textbf{n}} - \nu^*(t) \, {d}_{-\textbf{n}}^\dag  , \\ \label{BT01b}
{b}_{-\textbf{n}} &=&   \mu(t) \, {d}_{-\textbf{n}} - \nu^*(t) \, {c}_{\textbf{n}}^\dag ,
\eeq
where, $\mu \equiv \mu_n$ and $\nu \equiv \nu_n$, are given by
\beq\label{MU02}
\mu(t) &=& \frac{1}{2} \left( \sigma \sqrt{M \omega_n} +  \frac{1}{\sigma\sqrt{M \omega_n}}  - i\dot{\sigma} \sqrt{\frac{M}{\omega_n }}  \right) , \\ \label{NU02}
\nu(t) &=&\frac{1}{2} \left( \sigma \sqrt{M \omega_n} -  \frac{1}{\sigma\sqrt{M \omega_n}}  - i\dot{\sigma} \sqrt{\frac{M}{\omega_n }}  \right) ,
\eeq
with, $|\mu|^2 - |\nu|^2 = 1$ for all time.

Let us now compute the quantum state of the perturbation modes in one single universe of the entangled pair. From (\ref{VS01}) the composite state of the perturbation modes in the two entangled universes can be written in the density matrix formalism as
\be\label{RHO01}
\rho = | {0}_a {0}_b \rangle \langle {0}_a {0}_b | .
\ee
Using the Bogolyubov transformation (\ref{BT01a}-\ref{BT01b}) the vacuum state of the invariant representation, $ | {0}_a {0}_b \rangle$, can be written as \cite{Mukhanov2007}
\be
| {0}_a {0}_b \rangle = \prod_\textbf{n} \frac{1}{|\mu|} \left( \sum_{N=0}^\infty \left( \frac{\nu}{\mu} \right)^N | N_{c,\textbf{n}} N_{d,-\textbf{n}}\rangle \right) ,
\ee
where, 
\be
|N_{c, \textbf{n}}\rangle = \frac{(c_\textbf{n}^\dag)^N}{\sqrt{N!}} |0_{c,\textbf{n}}\rangle \ , \ |N_{d,- \textbf{n}}\rangle = \frac{(d_{-\textbf{n}}^\dag)^N}{\sqrt{N!}} |0_{d,-\textbf{n}}\rangle ,
\ee
are the number states of the diagonal representation (\ref{DR01a}-\ref{DR01b}). It means that the vacuum state of the invariant representation is full of particles of the scalar field and gravitons of the gravitational field. The number and properties of the particles depend on the parameters $\mu$ and $\nu$ and, thus, they depend on the rate of entanglement between the universes. The effects of the entanglement between the two universes can thus be indirectly observed because they are encoded in the observable state of the perturbation modes. These effects have no classical analogue so they should entail distinguishable features of the inter-universal entanglement and of the whole multiverse proposal too.

\section{Quantum thermodynamics of the perturbation modes}

Let us consider the quantum state of the particles and gravitons in just one single universe of the entangled pair. The reduced density matrix that represents the quantum state of the particles in one single universe alone can be obtained by tracing out from the density matrix (\ref{RHO01}) the state of the particles in the partner universe. It  yields \cite{RP2010, RP2017a, RP2017b}
\be\label{RHO02}
\rho_c = \text{Tr}_d {\rho}= \prod_\textbf{n} \frac{1}{Z_n} \sum_N e^{-\frac{1}{T_n} (N+\frac{1}{2})} | N_{c,\textbf{n}} \rangle \langle N_{c,\textbf{n}}| ,
\ee
where, $Z_n^{-1} = 2 \sinh\frac{1}{2T_n}$, and
\be\label{TE01}
T_n \equiv T_n(t) = \frac{1}{ \ln \frac{|\mu(t)|^2}{|\nu(t)|^2}} = \frac{1}{\ln\left( 1 + |\nu(t)|^{-2}\right)} .
\ee
The density matrix $\rho_c$ represents a quasi-thermal state whose thermal properties depend on the rate of entanglement between the universes. In particular, the specific temperature of entanglement (\ref{TE01}) is a measure of the entanglement \cite{RP2017a} between the field particles and the gravitons of the two entangled universes. The largest modes of the particles and gravitons do not feel the effect of the entanglement  because for large modes, $\nu \rightarrow 0$ and $T\rightarrow 0$, so these modes are in the vacuum states. However, for the shorter modes the effects of the inter-universal entanglement may be significant.

The modes are not really thermalized until the temperature $T_n$ becomes the same for all modes. Even though, one can define, for each mode, the thermodynamical magnitudes of entanglement associated to the thermal state (\ref{RHO02}). They are given by \cite{RP2011b}
\begin{eqnarray}\label{eq631}
E(a) &=& {\rm Tr}\left( \hat{\rho}_c(a) \hat{H}(a) \right) , \\  \label{eq632}
Q(a) &=& \int^a {\rm Tr}\left( \frac{d \hat{\rho}_c(a')}{d a'} \hat{H}(a') \right) da' , \\ \label{eq633}
W(a) &=& \int^a {\rm Tr}\left( \hat{\rho}_c(a') \frac{d \hat{H}(a')}{d a'}  \right) da' ,
\end{eqnarray}
where ${\rm Tr}(\hat{O})$ stands for the trace of the operator $\hat{O}$, and $H$ is the Hamiltonian of the harmonic oscillator (\ref{Hn}). For the density matrix $\rho_c$ in (\ref{RHO02}) it yields \cite{RP2011b, RP2017c}
\beq\label{Eent}
E_n(t) &=& \frac{\omega_n}{2} \cotanh\frac{1}{2 T_n} = \omega_n \left( N_n +\frac{1}{2}\right) , \\
Q_n(t) &=& \frac{\omega_n}{2} \cotanh\frac{1}{2 T_n} - \omega_n T_n \ln\sinh\frac{1}{2 T_n}, \\
W_n(t) &=& \omega_k T_n \ln\sinh\frac{1}{2 T_n} ,
\eeq
where, $N_n \equiv |\nu|^2$. The first principle of thermodynamics, $d E_n(t) = \delta Q_n(t) + \delta W_n(t)$, is satisfied for all modes $\textbf{n}$ individually, and the energy densities that correspond to $E_n$, $Q_n$, and $W_n$, are given by
\be\label{edensity}
\varepsilon_n = \frac{E_n}{V} \ , \ q_n = \frac{Q_n}{V} \ , \ w_n = \frac{W_n}{V} ,
\ee
with, $V = a^3(t)$, being the volume of the space.  The entropy of entanglement \cite{Horodecki2009, RP2011b} can also be easily obtained from the von Neumann formula
\be
S(\rho_c) = - \rm{Tr}\left( \rho_c \ln \rho_c \right) ,
\ee
with $\rho_c$ given by (\ref{RHO02}). It yields \cite{RP2012}
\begin{equation}
\label{eq68}
S_\text{ent}(a) = |\mu|^2 \, \ln |\mu|^2 - |\nu|^2 \, \ln |\nu|^2 ,
\end{equation}
from which it can be checked that the second principle of thermodynamics is also satisfied \cite{RP2011b}.

It is worth noticing that the energy of the vacuum state of the invariant representation is the same as the energy of the thermal state (\ref{RHO02}) of the diagonal representation. The former is given by 
\be
E_0^I =  _I\langle 0 | H | 0\rangle_I .
\ee
It yields \cite{Leach1983}
\be\label{EI}
E^I_0 = \frac{\omega_n}{4} \left( \sigma^2 M \omega_n + \frac{1}{\sigma^2 M \omega_n} + \frac{M \dot{\sigma}^2}{\omega_n} \right) .
\ee
On the other hand, the energy of the thermal state (\ref{RHO02}) in the diagonal representation is given by (\ref{Eent}) with, $N_n = |\nu|^2$, which can be also written as
\be\label{ED}
E^D_{th} = \frac{\omega_n}{2} \left( |\mu|^2 + |\nu|^2 \right) .
\ee 
By using the values of  $\mu$ and $\nu$ given in (\ref{MU02}-\ref{NU02}) it can be checked that (\ref{EI}) and (\ref{ED}) yield the same value. The energy is therefore conserved, as it was expected and the thermal state $\rho_c$ in (\ref{RHO02}) entails just a redistribution of the modes with the same total energy.

\section{Observable imprints}

Three potentially observable effects are expected to be caused by the creation of universes in entangled pairs. First, the boundary condition imposed on the state of the universes may modify the effective value of their Friedmann equation, given initially by (\ref{FE04}). Let us notice that the Wheeler-DeWitt equation (\ref{WDW02}) can be formally considered as the generalized equation of a harmonic oscillator with time dependent mass, $\mathcal{M}(a) = a$, and frequency $\Omega_K(a)$ given by (\ref{OMEGAK}), with the scale factor formally playing the role of the time variable. The quantization of that generalized harmonic oscillator is the basis of the so-called third quantization formalism. In that context, a similar argument to that applied to the perturbation modes can be given for the states of the wave function $\Psi_K$. We can then impose that the quantum state of the homogeneous and isotropic background should be the vacuum state of the invariant representation associated to the generalized harmonic oscillator (\ref{WDW02}). The vacuum state of the invariant representation, which is a stable ground state along the entire evolution in the minisuperspace, turns out to be full of entangled pairs of universes in the diagonal representation \cite{RP2017c}. The expected value of the generalized Hamiltonian $\mathcal{H}$ of the harmonic oscillator (\ref{WDW02}) is given, similarly to (\ref{EI}),  by
\beq\nn
\langle 0 |\mathcal{H}|0\rangle &=& \frac{\Omega_K}{4} \left( R^2 \mathcal{M}\Omega_K + \frac{1}{R^2 \mathcal{M}\Omega_K } + \frac{\mathcal{M}}{\Omega_K} \left(\frac{dR}{da}\right)^2 \right) \\
&\equiv& \frac{\tilde{\Omega}_K}{2} .
\eeq
Then, the evolution of the entangled universes is effectively determined  by a modified Friedmann equation given now, instead of by (\ref{FE04}), by
\be\label{FE05}
\frac{d a}{d t} = \pm \frac{\tilde{\Omega}_K}{a} .
\ee
In the adiabatic limit, for a large value of the scale factor,
\be
R \approx \frac{1}{\sqrt{\mathcal{M} \Omega_K}} .
\ee
Then,
\be
\tilde{\Omega}_K \approx \Omega_K \left( 1 + \frac{1}{8 \Omega_K^2} \left( \frac{d}{d a} \log(\mathcal{M}\Omega_K)\right)^2 \right) .
\ee
With the values, $\mathcal{M}=a$ and $\Omega_K \approx H(\phi) a^2$, it is obtained
\be
\tilde{\Omega}_K \approx H a^2 + \frac{9}{8 H a^4} .
\ee
The modified Friedmann equation (\ref{FE05}) yields
\be\label{SCF02}
a(t) = a_0 \left( e^{6 H \Delta t} -  1 \right)^\frac{1}{6} .
\ee
At late times the scale factor (\ref{SCF02}) evolves in an exponential way. However, the entanglement between the universes produce a pre-inflationary stage that should leave observable consequences \cite{Scardigli2011, Bouhmadi2011}.

Another modification of the Friedmann equation that is expected to leave observable imprints in the properties of our universe is the backreaction of the perturbation modes, given by the energy density associated to the energy (\ref{EI}) (or (\ref{ED})). In the case of an exactly flat DeSitter expansion the value of $\sigma$ that satisfies the boundary condition (\ref{SIGMBC01}) is given by (\ref{SIGM02}) with $\sigma_1$ and $\sigma_2$ given by
\beq\label{SIG1}
\sigma_1 &=& \sqrt{\frac{\pi}{2 H}} a^{-\frac{3}{2}} \mathcal{J}_q(\frac{n}{H a}) , \\ \label{SIG2}
\sigma_2 &=& \sqrt{\frac{\pi}{2 H}} a^{-\frac{3}{2}} \mathcal{Y}_q(\frac{n}{H a}) ,
\eeq
where $\mathcal{J}_q(x)$ and $\mathcal{Y}_q(x)$ are the Bessel functions of first and second kind and order, $q = \sqrt{\frac{9}{4}- \frac{m^2}{H^2}}$. For the case, $m \ll H$ ($q \approx \frac{3}{2}$), $\sigma$ turns out to be
\be\label{SIGM05}
\sigma^2 \approx \frac{H^2 a^2 + n^2}{a^2 n^3} ,
\ee
and the energy (\ref{EI}) yields
\be\label{En01}
E_n = \frac{H x}{2} \left( 1 + \frac{1}{2} (1 + \frac{m^2}{H^2}) x^{-2} + 2 \frac{m^2}{H^2} x^{-4}   \right) ,
\ee
with, 
\be
x \equiv \frac{n }{H a} = \frac{n_\text{ph}}{H} \sim \frac{H^{-1}}{L_\text{ph}} ,
\ee
where, $L_\text{ph}$, is the physical wave length and $H^{-1}$ is the distance to the Hubble horizon. The problem now is that the energy of the backreaction (\ref{En01}) turns out to be divergent when it is summed over all modes. A cut-off has to be imposed. Following Refs. \cite{Mersini2008c, Mersini2008d}, the energy of the modes can be integrated from the value $n=ab$, where $b$ is the SUSY breaking scale of the subjacent landscape, to the value $n = a H$, disregarding thus the superhorizon modes. Then, it is obtained
\beq\label{ESH01}
\varepsilon &=& \frac{1}{V} \int_{ab}^{aH} dn \, n^2 E_n = \\ \nn
&=& \frac{H^4}{8} \left\{ 1 - \frac{m^2}{H^2} \log\frac{b^2}{H^2} + \left( 1+ \frac{m^2}{H^2} \right) \left( 1 - \frac{b^2}{H^2} \right) \right\} ,
\eeq
where terms of higher order have been disregarded. The energy shift (\ref{ESH01}) can be seen as a correction to the effective value of the potential of the scalar field, an effect that is expected to produce observable imprints in the properties of the CMB \cite{Mersini2017a, DiValentino2017a, DiValentino2017b}.

Finally, the third effect that is expected to leave observable imprints in the properties of the CMB is the spectrum of fluctuation of the thermal state (\ref{RHO02}) caused by the entanglement between the universes. The fluctuations of the perturbation modes can be obtained from
\be\label{FLUCT00}
\delta\phi_\textbf{n} = \frac{n^\frac{3}{2}}{2 \pi} \Delta\phi_n ,
\ee
where
\be
(\Delta\phi_n)^2 = \langle |\phi_n|^2 \rangle - |\langle \phi_n \rangle|^2 .
\ee
In the vacuum state of the invariant representation \cite{Leach1983}, $|\langle \phi_n \rangle| = 0$, and
 \be\label{FLUCT01}
 \langle |\phi_n|^2 \rangle = \frac{\sigma^2}{2} .
 \ee 
In the case of a DeSitter spacetime $\sigma_1$ and $\sigma_2$ in (\ref{SIG1}-\ref{SIG2}) are the real and imaginary parts of the Bunch-Davies vacuum, so $\sigma$ is essentially the modulus of the Bunch-Davies modes and, thus, the fluctuations of the invariant vacuum, given by (\ref{FLUCT00}) with (\ref{FLUCT01}), turn out to yield the customary expression \cite{Mukhanov2007}
\be
\delta\phi_\textbf{n} = \frac{H}{\sqrt{8\pi}} x^\frac{3}{2} \left( \mathcal{J}_q^2(x) + \mathcal{Y}_q^2(x) \right)^\frac{1}{2} .
\ee
However, the creation of universes in entangled pairs induces the perturbation modes to be in the thermal state (\ref{RHO02}) of the diagonal representation and then,
 \beq
 \langle |\phi_n|^2 \rangle &=& \frac{1}{M \omega_n}  (|\nu|^2 + \frac{1}{2})  \\
 &=& \frac{1}{4 M \omega_n} \left( \sigma^2 M \omega_n + \frac{1}{\sigma^2 M \omega_n} + \frac{M \dot{\sigma}^2}{\omega_n} \right) .
 \eeq
 With the value of $\sigma$ given by (\ref{SIGM05}), it turns out that
 \be\label{QF01}
 \frac{\delta\phi_\textbf{n}^{th}}{\delta\phi_\textbf{n}^{I}} = \sqrt{\frac{1}{2}\left( 1 + \frac{x^2}{(1+x^2 )(1+\frac{m^2}{H^2 x^2})} \right) }  .
\ee
The  large modes ($x\gg 1$) are in the vacuum state and then, $\delta\phi_\textbf{n}^{th} \approx \delta\phi_\textbf{n}^{I}$. However, the departure is significant for the horizon modes, $x \sim 1$. This is a distinctive effect of the creation of the universes in entangled pairs that should leave an observable imprint in the properties of the CMB.  It has no analogue in the context of an isolated universe so therefore, it entails a distinguishable effect of the whole multiverse proposal.


\section{Summary and conclusions}

We have presented a detailed model of a multiverse made up of pairs of universes whose quantum mechanical states are entangled. The existence of the multiverse, although bizarre at first sight, is something that has been implicitly considered from the very beginning of quantum cosmology. Each semiclassical branch of the general solution of the Wheeler-DeWitt equation represents a spacetime background with matter fields propagating therein, i.e. it represents a different realization of the universe. The customary approach has generally consisted of considering one of these branches as the representative of our universe and disregard the rest of them because it seems meaningless to physically consider external elements to the universes. However, we have shown that quantum correlations and other non-local interactions may exist between the states of the universes and that they may leave observable imprints in the properties of a universe like ours. In particular, we have shown that the creation of universes in entangled pairs correlate the quantum states of the matter fields that propagate in their respective spacetimes.

We have shown as well that the most natural way in which the universes can be created is in entangled pairs that conserve the momentum conjugated to the scale factor, in a parallel way as particles are created in entangled pairs that conserve the total momentum in a quantum field theory. The momentum conjugated to the scale factor, however, depends on the expansion rate of the universes. Thus, the opposite values of the momentum in the pair of universes is related to the opposite expansion rates of the universes in terms of a common time variable. Nevertheless, the time experienced by the internal observers in their particle physics experiments, i.e. the time measured by actual clocks, are related by an antipodal like symmetry, $t_I = - t_{II}$. Then, from the point of view of the internal observers the universes are both expanding universes.

The quantum states of the particles of the matter fields and the gravitons of the spacetime that propagate in the two entangled universes become entangled too. The most appropriate boundary condition seems to be that the fields are in the composite vacuum state of the invariant representation. This is a stable representation of the vacuum state along the entire evolution of the fields. However, in terms of the instantaneous diagonal representation of the corresponding Hamiltonian, which would represent the state of the particles measure by internal observers, the quantum state of the field turns out to be given by a quasi thermal state whose thermodynamical magnitudes of entanglement depend on the rate of entanglement between the universes. Thus, the inter-universal properties of entanglement may be encoded in the quantum state of the matter fields that propagate in our universe.

We expect three observable effects caused by the creation of universes in entangled pairs. The first one would be caused by the entanglement of the background spacetimes of the universes. It would modify the effective value of the Friedmann equation by introducing a pre-inflationary stage in the evolution of the universe that might leave observable imprints in the properties of the CMB provided that inflation does not last for too long. A second effect would be caused by the backreaction of the inhomogeneous degrees of freedom. It would entail a modification of the effective value of the potential of the scalar field that would have a direct consequence in the properties of the inflationary expansion and thus, in the observed properties of the early universe. Finally, the spectrum of fluctuations of the perturbation modes for the thermal state induced by the entanglement of the partner universes, is significantly different from the one expected from an unentangled universe. This is then a distinctive feature of the creation of the universes in entangled pairs that has no analogue in the context of an isolated universe. It is therefore a distinguishable feature of the whole multiverse proposal.

We have shown therefore that the multiverse is a testable proposal. The process that my happen in the multiverse would eventually leave their imprints in the observable properties of the single universes and, thus, they become testable. The door is now open for the study of a wide variety of new cosmic phenomena. Let us notice that these effects are expected to be residual effects of the underlying theory, whether this is one of the string theories of the quantum theory of gravity. Thus, they may help us to test these most fundamental theories.




\bibliographystyle{apsrev4-1}

%

\end{document}